\documentstyle[aps]{revtex}

\newcommand{\beq}{\begin{equation}}
\newcommand{\eeq}{\end{equation}}
\newcommand{\four} {  {}^{(4)}\kern-1pt  }
\newcommand{\ben}{\begin{eqnarray}}
\newcommand{\een}{\end{eqnarray}}
\newcommand{\f}{\frac}

\catcode`@=11     
\def\eqalign#1{\null\,\vcenter{\openup\jot\m@th
  \ialign{\strut\hfil$\displaystyle{##}$&$\displaystyle{{}##}$\hfil
      \crcr#1\crcr}}\,}
\catcode`@=12   

\begin{document}

\title{\begin{flushright}
hep-th/0107039 \\
$~$ \\
\end{flushright}
DEFORMED BOOST TRANSFORMATIONS THAT\\ SATURATE
AT THE PLANCK SCALE}

\author{{\bf N.R.~Bruno} and {\bf G.~Amelino-Camelia}}
\address{Dipart.~Fisica,                        
Univ.~Roma ``La Sapienza'', and INFN Sez.~Roma1\\   
P.le Moro 2, 00185 Roma, Italy}                     

\author{{\bf J.~Kowalski-Glikman}} 
\address{Institute for Theoretical Physics, 
University of Wroc{\l}aw,\\                     
pl. Maxa Borna 9, 50--204 Wroc{\l}aw, Poland}   

\maketitle

\begin{abstract}
We derive finite boost transformations based on the Lorentz sector
of the bicross-product-basis $\kappa$-Poincar\'{e} 
Hopf albegra. We emphasize the role of these boost transformations
in a recently-proposed new relativistic theory.
We find that when the (dimensionful) deformation parameter is identified 
with the Planck length, which together with the speed-of-light constant
has the status of observer-independent scale in the new relativistic
theory, the deformed boosts saturate at the value of momentum
that corresponds to the inverse of the Planck length.
\end{abstract}

\section{Introduction}

Several approaches to the unification of General Relativity and Quantum Mechanics
have led to arguments in favour of the emergence of a minimum length
and/or a maximum momentum, possibly connected with the Planck length
($L_p \sim 10^{-33} cm$) and this has also motivated some authors
(see, {\it e.g.}, Refs.~\cite{satusussk,satugaray})
to argue that somehow boosts should saturate at the Planck scale.
A framework that can be used to implement this type of concepts
in consistent relativistic theories
was recently proposed in Ref.~\cite{gacdsr}, where,
following a close analogy with our 
present description of fundamental physics (in which
Galileo's Relativity Principle coexists with the observer-independent scale $c$,
the speed-of-light fundamental constant), it was shown 
that logically consistent relativistic theories can host both $c$
and a second scale, possibly connected with 
the Planck length, as observer-independent scales.
One example of this new type of Relativity postulates, 
an example based on the introduction
of a deformed dispersion relation, was analyzed in detail~\cite{gacdsr} 
obtaining various results in leading order in the second (Planck-scale
related) observer-independent scale, including the nature of the transformation
rules between different inertial observers
and kinematic rules\footnote{Additional work on these kinematic
rules was then reported in Ref.~\cite{gacmarz}.} for particle-production
processes, emphasizing those aspects that lead to the emergence of
a minimum length, a minimum length uncertainty and a maximum 
momentum (minimum wavelength). 
In particular, it was found 
(again within a leading-order analysis~\cite{gacdsr})
that consistency with the new postulates 
requires that (infinitesimal and finite) transformations 
between different
inertial observers be described in terms of the generators of one
of the $\kappa$-Poincar\'{e} 
Hopf albegras~\cite{lunoruto,maru,luruto,luruza,luno},
and that the action of boosts saturates at momenta corresponding to 
the second observer-independent scale.
This connection with $\kappa$-Poincar\'{e} was then exploited 
in Ref.~\cite{jkgminl} as a guiding principle 
for obtaining exact (all-order) 
results on the maximum,
observer-independent, momentum scale, and on
some asymptotic (infinite-energy)
features of the new relativistic theory.

In this paper we intend to further exploit the connection 
with $\kappa$-Poincar\'{e}
to obtain the exact (all-order) form of the finite boost transformations,
generalizing one of the leading-order results of Ref.~\cite{gacdsr}.
The relevant properties of operators obtained by exponentiation of
the generators of the $\kappa$-Poincar\'{e} algebra here of interest
were not previously analyzed, but for another example 
of $\kappa$-Poincar\'{e} algebra a similar analysis was attempted~\cite{ruquasi},
encountering several difficulties (in particular, the exponentiatied
action of the generators could only be described through implicit formulas).
We shall show that no such difficulties is encountered
in the analysis of the $\kappa$-Poincar\'{e} algebra here of interest.

As additional motivation for the analysis reported in the following
Sections we observe that, in addition to their significance
for the logical consistency of the new type of relativistic theories, 
the possibility of deformed boosts has recently attracted interest
also as a tool to address certain problems in phenomenology.
In particular, puzzling observations of ultra-high-energy
cosmic rays and multi-TeV photons from Markarian 501 have been analyzed
as an indication of boost deformation~\cite{kifu,aus,ita,gactp}.
Also some cosmological scenarios based on $\kappa$-Poincar\'{e}
were analyzed, leading to encouraging preliminary results~\cite{jkgcosm,stjo}.

\section{Finite boost transformations}

The $\kappa$-Poincar\'{e} Hopf algebra that is relevant for the 
example of new relativistic theory studied in Refs.~\cite{gacdsr,jkgminl}
is the one proposed by Majid and Ruegg in Ref.~\cite{maru} (the
so-called ``bicrossproduct basis").
For our analysis of finite boosts only the algebra sector of this
Hopf algebra is relevant, and we note the commutation relations
here for completeness:\footnote{Since the second observer-independent
scale appears several times in several of our formulas we find 
convenient to denote it with $\lambda$ rather than the more cumbersome
notation ${\tilde L}_p$ adopted in Ref.~\cite{gacdsr}. We of course
maintain the intuition that this dimensionful parameter should 
be naturally identified with the Planck length, up to a numerical factor
of order 1 and a possible sign difference. Also notice that in the
$\kappa$-Poincar\'{e} literature the mass parameter $\kappa$ is used instead
instead of the length parameter $\lambda$ ($\kappa \equiv 1/\lambda$).}
\begin{eqnarray} \label{1}
[M_{\mu\nu},M_{\rho\tau}]
=&&
\displaystyle
i\left(\eta_{\mu\tau}M_{\nu\rho} -
\eta_{\nu\rho}M_{\nu\tau} +\eta_{\nu\rho}M_{\mu\tau}
  -\  \eta_{\nu\tau}M_{\mu\rho} \right),
\cr \cr
 [M_{i},k_{j}] =&&  i\epsilon_{ijk}k_{k}\, ,
\quad [M_{i},\omega]= 0\, ,
\cr \cr
\displaystyle
  [N_{i}, k_{j}] = && i\delta_{ij}
 \left( {1\over {2\lambda}} \left(
 1 -e^{-2{\omega\lambda}}
\right) + {\lambda\over 2} \vec{k}\, ^{ 2}\, \right)
- \ {i\lambda} k_{i}k_{j} ,
\cr \cr
 \left[N_{i},\omega\right] = && ik_{i}\, ,
\cr \cr
\displaystyle
  [P_{\mu},P_{\nu}] = && 0\, ,
\end{eqnarray}
where $P_\mu =(\omega, k_i)$ are the time and space components of the
four-momentum generators and $M_{\mu\nu}$ are modified Lorentz generators
with rotations $M_k = \frac12\epsilon_{ijk}M_{ij} $ and boosts
$N_i = M_{0i}$.
Also important for our analysis is
the dispersion relation that follows from
these $\kappa$-Poincar\'{e} algebraic relations
\beq\label{9}
\f{e^{\lambda\omega}+e^{-\lambda\omega}-2}{\lambda^2}
-\vec{k}^2e^{\lambda\omega}=m^2 ~,
\eeq   
which of course corresponds to a Casimir of (\ref{1})
(just as the special-relativistic dispersion relation
corresponds to a Casimir of the standard Lorentz algebra).
The eigenvalues $m^2$ of the $\kappa$-deformed Casimir are 
related to the physical mass $M$ (rest energy) by 
\beq\label{9bis}
\f{e^{\lambda M}+e^{-\lambda M}-2}{\lambda^2}=m^2 ~.
\eeq   

Clearly according to (\ref{1}) the action of rotations is completely
conventional (undeformed).
The action of the boosts $N_i$ needs instead a deformation,
to reflect the properties of the $\lambda$-dependent 
commutator $[N_{i}, k_{j}]$.
As announced, we intend to construct finite deformed boosts transformations.
This can be done in close analogy with the corresponding 
analysis of ordinary Lorentz boosts. 
A particle which,
for a given first observer, has four-momentum $(\omega^0,\vec{k}^0)$
will have four-momentum $(\omega,\vec{k})$
for a second observer
in relative motion, with boost/rapidity parameter $\xi$,
with respect to the first observer.
The starting point for obtaining
the relation between $(\omega,\vec{k})$ and $(\omega^0,\vec{k}^0)$
is the differential 
representation of the deformed boost generators.
From the algebra (\ref{1}) one can easily derive
this differential representation (of course, without loss of 
generality we can choose to focus on the boost that acts 
along the axis $1$)
\beq\label{2}
N_1=ik_1\frac{\partial}{\partial \omega}+i\left(\frac{\lambda}{2}\vec
k^2+\frac{1-e^{-2\lambda\omega}}{2\lambda}\right ) \frac{\partial}{\partial
k_1}-i\lambda k_1 \left(k_j\frac{\partial}{\partial k_j} \right ),
\eeq
and accordingly\footnote{Here we are making the implicit 
assumption that the action of the adjoint representation
be described by exponentiation of the generators.
This assumption is in fact justified~\cite{luruza}
in the specific $\kappa$-Poincar\'{e} Hopf algebra here of interest,
differently from the case of the Lorentz sector of other
$\kappa$-Poincar\'{e} Hopf algebra~\cite{luruza,ruquasi}.} 
the differential equations to be satisfied 
by $(\omega,\vec{k})$ are
\beq\label{3}
\frac{d}{d \xi} k_1 ({\xi})+\f{{\lambda}}{2}(k_1^2({\xi})-k_2^2 ( {\xi})
-k_3^2( {\xi}))+\f{e^{-2{\lambda}{\omega}(\xi)}-1}{2{\lambda}}=0
\eeq
\beq\label{4}
\f{d}{d{\xi}}{\omega}({\xi})-k_1({\xi})=0
\eeq
\beq\label{5}
\f{d}{d{\xi}}k_2({\xi})+{\lambda} k_1({\xi})k_2({\xi})=0
\eeq
\beq\label{6}
\f{d}{d{\xi}}k_3({\xi})+{\lambda} k_1({\xi})k_3({\xi})=0
\eeq

Differentiating (\ref{3}) and making use of the other equations
one obtains a non-linear second-order equation for $k_1({\xi})$: 
\beq\label{7}
{\frac {d^{2}}{d{{\xi}}^{2}}}k_1({\xi})+3\,\lambda\,k_1({\xi})
{\frac {d}{d{\xi}}}k_1({\xi})+{
\lambda}^{2}k_1^3({\xi}) -k_1({\xi})=0 ~.
\eeq
We find that the solutions of this equation are of the form
\beq\label{8}
k_1(\xi)=-\f{B}{\lambda}
\f{\cosh ({\xi}+\beta)}{(1-B\sinh ({\xi}+\beta))}\label{k11}
\eeq
where $B$ and $\beta$ are integration constants.

Corresponding solutions for the other components of the
four-momentum can be obtained by substituting Eq.~(\ref{8}) 
in Eqs.~(\ref{4}--\ref{6}). 
Our problem is therefore already reduced to the identification
of a few integration constants.


We determine the integration constants $B$, $\beta$ 
by imposing the obvious requirement
that $(\omega,\vec{k})=(\omega^0,\vec{k}^0)$
for $\xi=0$.
From (\ref{8}) one obtains
\beq\label{jj1}
k_1({\xi}=0)=k_1^0=-\f{B}{\lambda}\f{\cosh (\beta)}{(1-B\sinh (\beta))}
\eeq
and thus
\beq\label{jj2}
B=-\f{\lambda k_1^0}{\cosh(\beta)-\lambda k_1^0 \sinh(\beta)}
\eeq
Using this condition and introducing $A=\tanh(\beta)$
(simply replacing the unknown $\beta$ with the corresponding unknown $A$)
we can obtain from (\ref{8}) an expression of $k_1({\xi})$ with only 
one unknown:
\beq
k_1({\xi})
=k_1^0 \f{\cosh({\xi})+A\sinh({\xi})}{1
-\lambda k_1^0(A-A\cosh({\xi})-\sinh({\xi}))}  ~.
\label{10}
\eeq
Corresponding expressions for $\omega(\xi)$,$k_2(\xi)$,$k_3(\xi)$
are easily obtained from Eqs.~(\ref{4})--(\ref{6}):
\beq\label{12}
\omega(\xi)=\omega^0-\f{1}{\lambda}
\mbox{ln}\f{1}{1
-\lambda k_1^0(A-A\it{\cosh}(\xi)-\it{\sinh}(\xi))}
\eeq
\beq\label{13}
k_2(\xi)=\f{k_2^0}{1-\lambda k_1^0(A-A\it{\cosh}(\xi)-\it{\sinh}(\xi))}
\eeq
\beq\label{14}
k_3(\xi)=\f{k_3^0}{1-\lambda k_1^0(A-A\it{\cosh}(\xi)-\it{\sinh}(\xi))}
\eeq
We are therefore left with the task of expressing $A$ in terms
of $\omega^0$,$k_1^0$,$k_2^0$,$k_3^0$.
The sought condition is easily obtained combining
(\ref{3}) and (\ref{10})
\beq\label{jjj7}
k_1^0A-\lambda{k_1^0}^2 = \left.{\frac {d}{d{\xi}}}k_1({\xi})\right|_{{\xi}=0}
=-\f{{\lambda}}{2}
({k_1^0}^2-{k_2^0}^2-{k_3^0}^2)
+\f{1 - e^{-2{\lambda}{\omega^0}}}{2{\lambda}} ~,
\eeq
from which it follows that 
\beq\label{11}
A=\sinh(\lambda\omega^0)\f{e^{-\lambda\omega^0}}{\lambda k_1^0}
+\f{\lambda}{2k_1^0}\vec{k_0}^2
\eeq

The equations (\ref{10}), (\ref{12}), (\ref{13}) and (\ref{14}),
with $A$ expressed in terms
of $\omega^0$,$k_1^0$,$k_2^0$,$k_3^0$ through (\ref{11}),
describe the exact $\kappa$-deformed boost transformations.
It is easy to verify that 
they satisfy\footnote{In the new relativistic theory the deformed 
dispersion relation acquires~\cite{gacdsr} the status of an 
observer-independent property, and it must therefore 
be an invariant of boost transformations. At the mathematical
level this is assured~\cite{gacdsr,jkgminl} by the fact that 
the deformed dispersion relation corresponds, as mentioned, to
a Casimir of (\ref{1}).} 
the dispersion relation (\ref{9}). 
It is also easy to verify that,
of course, in the $\lambda \rightarrow 0$ limit our transformation
rules reduce to ordinary Lorentz boost transformations,
and the leading order in $\lambda$
reproduces the corresponding result obtained in Ref.~\cite{gacdsr}.

\section{Range of the boost parameter and maximum momentum}

For ordinary Lorentz boosts the $\xi$ 
parameter can take any real value. In this Section we show that
the same property holds for our deformed boosts for $\lambda >0$,
while for $\lambda <0$ the $\xi$ 
parameter can vary only within a finite range.
We also show that for $\lambda >0$
our deformed boosts saturate at a maximum value 
of momentum: $|\vec{k}| =1/\lambda$.

In preparation for the study of the range of $\xi$,
let us start by analyzing some relevant properties of the 
integration constant $A$.
For simplicity let us focus on the case $k_2^{0} = k_3^{0}=0$ 
and let us denote $k_1^{0}$ simply 
by $k^{0}$; then $A$ takes the form
\beq\label{jj11}
A=\sinh(\lambda\omega^{0})\f{e^{-\lambda\omega^{0}}}{\lambda k^{0}}
+\f{\lambda}{2}k^{0}
~.
\eeq
Using the deformed dispersion relation (\ref{9}), which allows to express
$k^{0}$ as a function of $\omega^{0}$
\beq\label{jj22}
k^{0}={\pm}\f{\sqrt{1+e^{-2\lambda\omega^{0}}-(2+m^2\lambda^2)
e^{-\lambda\omega^{0}}}}{|\lambda|}
~,
\eeq
and using (\ref{9bis})
to express the Casimir eigenvalues $m^2$
in terms of the physical mass $M$,
we obtain a useful formula for $A$
\beq\label{19}
A= \mbox{sign}(k)\, 
\mbox{sign}(\lambda)\f{1- \cosh(\lambda M)x}{\sqrt{1+x^2-2\cosh(\lambda M)x}}
\eeq
where we have also introduced $x \equiv e^{-\lambda \omega^0}$. 

For $\lambda > 0$, $x\in (0, e^{-\lambda M}]$
and from (\ref{19}) it follows that for positive $k$  
the value of $A$ varies from $1$ to $+\infty$
as $x$ varies from $0$ to $e^{- \lambda M}$, while for negative $k$
the value of $A$ varies 
from $-1$ to $-\infty$
as $x$ varies from $0$ to $e^{- \lambda M}$.

For $\lambda <0$, $x\in [e^{|\lambda| M},+\infty)$
and from (\ref{19}) it follows that for positive $k$  
the value of $A$ varies from $+\infty$ to $1$
as $x$ varies from $e^{|\lambda| M}$ to $+\infty$, while for negative $k$
the value of $A$ varies 
from $-\infty$ to $-1$ 
as $x$ varies from $e^{|\lambda| M}$ to $+\infty$. 

Using these properties of $A$ it is easy to establish
the range of allowed values of the boost parameter $\xi$
and to establish the main characteristics of the dependence
of momentum and energy on $\xi$.

For positive $\lambda$ it is possible to vary $\xi$ from $0$
toward both $+\infty$ and $-\infty$ without ever encountering
any singularities. (For positive $\lambda$
the denominator $1-\lambda k(A-A\it{\cosh}\xi-\it{\sinh}\xi) $
in (\ref{10}) and (\ref{12}) never vanishes.) 
All real values of $\xi$ are therefore allowed
for $\lambda >0$, just as in the case of
ordinary Lorentz boosts ($\lambda =0$).
The derivative $dk/d\xi$ vanishes only for one value 
of $\xi$ (a saddle point for $k(\xi)$)
and in that point the momentum vanishes and
the energy reaches its minimum
value $M$.
In the limit $\xi \rightarrow +\infty$ one 
finds $\omega \rightarrow +\infty$ while $k \rightarrow 1/\lambda$,
while in the limit $\xi \rightarrow -\infty$ one 
finds $\omega \rightarrow +\infty$ while $k \rightarrow - 1/\lambda$.
So energy is still 
unbounded from above, just like in the ordinary Lorentz-boost
case, while for positive $\lambda$ the new boost transformations
are such that momentum saturates at $|k| = 1/\lambda$.
Our result on the exact transfomation rules therefore extends 
the maximum-momentum
analysis reported in Ref.~\cite{gacdsr}, which was based on
the form of the transformation rules in leading order in $\lambda$
and found that for positive $\lambda$ the action of boosts
starts saturating as $|k|$ approaches $1/\lambda$
(the exact saturation result we obtained here
was of course not within the grasp of the leading-order analysis).
The maximum-momentum result that follows from our transformation
rules also reflects the property
of the $\kappa$-Poincar\'{e} dispersion relation (\ref{9}),
already emphasized in Ref.~\cite{jkgminl}, that for positive $\lambda$
connects the infinite-energy limit with the limit $|k| = 1/\lambda$.


The fact that the $\xi$ parameter can take any real value and that
the new observer-independent scale $\lambda$ acquires the intuitive
role of inverse of the maximum momentum renders the case $\lambda > 0$ 
rather attractive for applications in quantum-gravity research.

The situation is significantly different and
somewhat less intuitive in the case $\lambda < 0$.
For negative $\lambda$,
increasing $\xi$ from $0$ the energy already diverges at
\beq\label{31}
\xi_+=\ln \left(\,{\frac {1-\,\lambda\,k A+\,\sqrt {1
-2\,\lambda\,k A+{\lambda}^{2}{k }^{2}}}{-\lambda\,k A-\lambda\,k}}\right)~,
\eeq
and decreasing $\xi$ from $0$ the energy already 
diverges at
\beq\label{41}
\xi_-=\ln \left(\,{\frac {1-\,\lambda\,k A-\,\sqrt {1-2\,\lambda\,k A
+{\lambda}^{2}{k}^{2}}}{-\lambda\,k A-\lambda\,k}}\right) ~.
\eeq
(For negative $\lambda$
the denominator $1-\lambda k(A-A\it{\cosh}\xi-\it{\sinh}\xi) $
in (\ref{10}) and (\ref{12}) vanishes at $\xi=\xi_+$ and $\xi=\xi_-$.)

For the momentum one finds that
for negative $\lambda$
the derivative $dk/d\xi$ vanishes only for one value 
of $\xi$ (again, a saddle point for $k(\xi)$)
and in that point the momentum vanishes (and
the energy reaches its minimum
value $M$), then to the left and to the right of this saddle point
the function $k(\xi)$ approaches singular asymptotes at $\xi_-$
and $\xi_+$; in fact,
$k \rightarrow + \infty$ for $\xi \rightarrow \xi_+$, while
$k \rightarrow - \infty$ for $\xi \rightarrow \xi_-$.

The fact that it does not predict a maximum momentum and
that $\xi$ is confined to the range $\xi_- < \xi < \xi_+$
might render the case $\lambda < 0$ less attractive 
for physics application, but this is still a very early
in the development of the new relativistic theories~\cite{gacdsr}
and our intuition might be changed by future studies.

\section{Conclusions}

The interesting properties of the finite rules of transformation
between different inertial observers here obtained provide
additional insight in the new relativistic theory 
proposed in Ref.~\cite{gacdsr} and further developed
in Refs.~\cite{jkgminl} and~\cite{gacmarz}.
The case $\lambda > 0$ is particularly interesting
since it corresponds to boosts that saturate when the momentum
gets to $|k| = 1/\lambda$.
The fact that 
these deformed boosts act on small momenta ($|k| \ll 1/\lambda$)
in a way that is basically identical 
to the one of ordinary Lorentz boosts but 
are then able to saturate at momenta $1/\lambda$
could lead to interesting conceptual and phenomenological developments.
It is also reassuring (at least from the limited prospective we
presently have on these new relativistic theories)
that for this case $\lambda > 0$ the range of the boost parameter
is just the same as in ordinary Special Relativity.

Our results for negative $\lambda$ (no maximum momentum,
finite range of the boost parameter) appear to be less encouraging.
Although none of the results we obtained can be used to exclude
the case $\lambda < 0$ on physical grounds,
it appears reasonable to focus future studies of this new relativistic
framework on the case with positive $\lambda$.

Our analysis also changes the intuition that emerged in some previous 
studies, also based on $\kappa$-Poincar\'{e} mathematics. In particular,
in Ref.~\cite{ruquasi} the action of operators obtained by
exponentiation of the generators in the Lorentz sector
of another $\kappa$-Poincar\'{e} 
Hopf algebra was analyzed, encountering several difficulties
and finally obtaining a description of these actions
that could only be expressed very implicitly (through complicated
integrals). The fact that in the $\kappa$-Poincar\'{e} 
Hopf albegra here of interest we did not encounter them
might indicate that these difficulties 
are not a general characteristic of $\kappa$-Poincar\'{e}.
It is reasonable to conjecture that this significant difference between
the $\kappa$-Poincar\'{e} Hopf algebra here analyzed and the one analyzed
in Ref.~\cite{ruquasi} be due to the fact, already emphasized in
Ref.~\cite{luruza}, that, while indeed in the case we considered
the exponentiation of the generators does correspond to 
the action of the adjoint representation, in the case
considered in Ref.~\cite{luruza} by exponentiating the 
generators in the Lorentz sector one does not obtain
the action of the adjoint representation.
Another possible element for the understanding of this different
behaviour of different $\kappa$-Poincar\'{e} Hopf algebra appears
to be provided by the type of duality emphasized in Ref.~\cite{maru},
which is enjoyed by the example here considered, but is not present
in other $\kappa$-Poincar\'{e} Hopf algebras.

\vfil
\eject

\section*{Acknowledgements} NRB thanks the
International Center for Relativistic Astrophysics (ICRA) 
at the University of Rome ``La Sapienza" for hospitality during
part of this work.
GAC acknowledges conversations with
J.~Ng and H.~Van Dam and would like to thank the Theoretical Physics
Group of the University of North Carolina for hospitality during
part of this work. JKG thanks the Department of Physics 
of the University of Rome ``La Sapienza'' for hospitality during
part of this work. JKG research is partially supported by 
the KBN grant 5PO3B05620.



\begin{thebibliography}{99}

\bibitem{satusussk} L.~Susskind,
Phys.~Rev.~{\bf D49}, 6606 (1994).

\bibitem{satugaray} L.J.~Garay, 
Int.~J.~Mod.~Phys.~{\bf A10}, 145 (1995).

\bibitem{gacdsr}  G. Amelino-Camelia, gr-qc/0012051, 
Int. J. Mod. Phys. {\bf D} (2001) in press;
hep-th/0012238, Phys. Lett. {\bf B510}, 255 (2001);
gr-qc/0106004.

\bibitem{gacmarz} G.~Amelino-Camelia and M.~Arzano, {hep-th/0105120}.

\bibitem{lunoruto} J. Lukierski, A. Nowicki, H. Ruegg and V.N. Tolstoy,
Phys. Lett. {\bf B264}, 331 (1991).

\bibitem{maru} S. Majid and H. Ruegg, Phys. Lett. {\bf B334},
348 (1994).

\bibitem{luruto} J. Lukierski, H. Ruegg and V.N. Tolstoy,
Proceedings of XXX Karpacz School, February 1994, ``Quantum
Groups: Formalism and Applications", Eds. J. Lukierski,
Z. Popowicz and J. Sobczyk, Polish Scientific Publishers PWN,
p. 359 (1995).

\bibitem{luruza} J. Lukierski, H. Ruegg and W.J. Zakrzewski, Ann.
Phys. {\bf 243}, 90 (1995).

\bibitem{luno} J. Lukierski and A. Nowicki, Proceedings of
Quantum Group Symposium at Group 21, (July 1996, Goslar) Eds.
H.-D. Doebner and V.K. Dobrev, Heron Press, Sofia, 1997, p. 186.

\bibitem{jkgminl} J. Kowalski-Glikman, {hep-th/0102098}.

\bibitem{ruquasi} J.~Lukierski, H.~Ruegg and W.~Ruhl,
Phys.~Lett.~{\bf B313}, 357 (1993).

\bibitem{kifu} T.~Kifune,
Astrophys.~J.~Lett.~{\bf 518}, L21 (1999).

\bibitem{aus} R.J.~Protheroe and H.~Meyer,
Phys.~Lett.~{\bf B493}, 1 (2000).

\bibitem{ita}
R.~Aloisio, P.~Blasi, P.L.~Ghia and A.F.~Grillo,
Phys.~Rev.~{\bf D62}, 053010 (2000).

\bibitem{gactp}  G. Amelino-Camelia and T. Piran, hep-ph/0006210,
Phys.~Lett.~{\bf B497}, 265 (2001);
astro-ph/0008107, Phys.~Rev.~{\bf D} (2001) in press.

\bibitem{jkgcosm} J. Kowalski-Glikman, Phys. Lett. {\bf B499}, 1 (2001).

\bibitem{stjo} S. Alexander and J. Magueijo, {hep-th/0104093}.

\end{thebibliography}
\end{document}